\documentclass[
 aip,
 amsmath,amssymb,
 reprint
]{revtex4-1}

\usepackage{graphicx}
\usepackage{dcolumn}
\usepackage{bm}

\usepackage[utf8]{inputenc}
\usepackage[T1]{fontenc}
\usepackage{mathptmx}
\usepackage{etoolbox}

\makeatletter
\def\@email#1#2{
 \endgroup
 \patchcmd{\titleblock@produce}
  {\frontmatter@RRAPformat}
  {\frontmatter@RRAPformat{\produce@RRAP{*#1\href{mailto:#2}{#2}}}\frontmatter@RRAPformat}
  {}{}
}
\makeatother
\begin{document}

\preprint{AIP/123-QED}

\title[Optimizing Off-Axis Fields for Two-Axis Magnetometry with Point Defects]{Optimizing Off-Axis Fields for Two-Axis Magnetometry with Point Defects}

\author{N. M. Beaver}
\altaffiliation{These authors contributed equally}
\affiliation{Department of Physics, Northeastern University, Boston, MA, 02115, USA}

\author{N. Voce}
\altaffiliation{These authors contributed equally}
\affiliation{Department of Physics, Northeastern University, Boston, MA, 02115, USA}

\author{P. Meisenheimer}
\affiliation{Department of Materials Science and Engineering, University of California, Berkeley, CA, 94720, USA}

\author{R. Ramesh}
\affiliation{Department of Materials Science and NanoEngineering, Rice University, Houston, TX, 77251, USA}
\affiliation{Department of Physics and Astronomy, Rice University, Houston, TX, 77251, USA}
\affiliation{Department of Physics, University of California, Berkeley, CA, 94720, USA}

\author{P. Stevenson}
 \email{p.stevenson@northeastern.edu}
\affiliation{Department of Physics, Northeastern University, Boston, MA, 02115, USA}

\date{\today}

\begin{abstract}
Vector magnetometry is an essential tool in characterizing the distribution of currents and magnetization in a broad range of systems. Point defect sensors, like the nitrogen vacancy (NV) center in diamond, have demonstrated impressive sensitivity and spatial resolution for detecting these fields. Measuring the vector field at a single point in space using single defects, however, remains an outstanding challenge. We demonstrate that careful optimization of the static bias field can enable simultaneous measurement of multiple magnetic field components with enhanced sensitivity by leveraging the nonlinear Zeeman shift from transverse magnetic fields, realizing an improvement in transverse sensitivity from $>200\mu T/\sqrt{Hz}$ (no bias field) to $30\mu T/\sqrt{Hz}$. This work quantifies the trade-off between the increased frequency shift from second-order Zeeman effects with decreasing contrast as off-axis field components increase, demonstrating the measurement of multiple components of the magnetic field from an exemplar antiferromagnet with a complex magnetic texture.
\end{abstract}

\maketitle

The ability to detect and image magnetic fields is of great significance in a broad range of technologies and applications, enabling new approaches to characterizing microelectronics\cite{garsiThreedimensionalImagingIntegratedcircuit2024,turnerMagneticFieldFingerprinting2020} and providing new insight into novel material systems\cite{thielProbingMagnetism2D2019,stefanMultiangleReconstructionDomain2021,tetienneNatureDomainWalls2015}. In particular, the ability to resolve the vector nature of magnetic fields at the nanoscale is a powerful capability, allowing, for example, different topological features in magnetic materials to be distinguished\cite{tanRevealingEmergentMagnetic2024,dovzhenkoMagnetostaticTwistsRoomtemperature2018a} and electron flow to be quantitatively imaged\cite{turnerMagneticFieldFingerprinting2020,palmImagingSubmicroampereCurrents2022}. 

Optically-addressable spin qubits, such as the nitrogen vacancy (NV) center in diamond\cite{chatterjeeSemiconductorQubitsPractice2021}, are able to realize this sought-after vector magnetometry capability with outstanding sensitivity and spatial resolution. NV center ensembles have been used to demonstrate vector magnetometry with sensitivities down to $<9\,pT/\sqrt{Hz}$ \cite{wangPicoteslaMagnetometryMicrowave2022} and diffraction-limited spatial resolution, while single NV centers have been used to probe nanoscale volumes in either static or scanning-tip modalities\cite{thielProbingMagnetism2D2019,tetienneNatureDomainWalls2015}. Nanoscale vector magnetometry is, however, far from a solved problem. Typically, one of two approaches is taken; one approach utilizes an ensemble of NV centers\cite{chenCalibrationFreeVectorMagnetometry2020}, providing excellent sensitivity but ultimately limiting resolution to the diffraction limit of the optical system. Alternatively, spatial images of the magnetic field obtained with a scanning tip system can be used to reconstruct the vector field\cite{chengObservationMagneticDomain2022}, albeit at the cost of of substantially increased acquisition time and ambiguities in the reconstruction method\cite{broadwayImprovedCurrentDensity2020}. Alternative approaches utilizing the nitrogen nuclear spin have been demonstrated, though these methods come with additional experimental complexity\cite{qiuNuclearSpinAssisted2021a,liuNanoscaleVectorDc2019}. Characterizing the vector components of weak magnetic fields at a single defect remains an outstanding challenge.

Here, we demonstrate and analyze a straightforward experimental approach which enables simultaneous detection of multiple magnetic field components with single NV centers in a continuous-wave (CW) modality, and demonstrate the detection of fields down to the $10\mu T$ level in an antiferromagnetic sample. Exploiting the nonlinear response of the NV center Hamiltonian to fields transverse to the zero-field quantization axis, we engineer an order-of-magnitude improvement in sensitivity to small transverse fields with minimal degradation of the NV center's exceptional sensitivity to other magnetic field components. Our approach enables both the parallel and transverse magnetic fields to be detected in a single measurement. 

\begin{figure}
\includegraphics[width=0.45\textwidth]{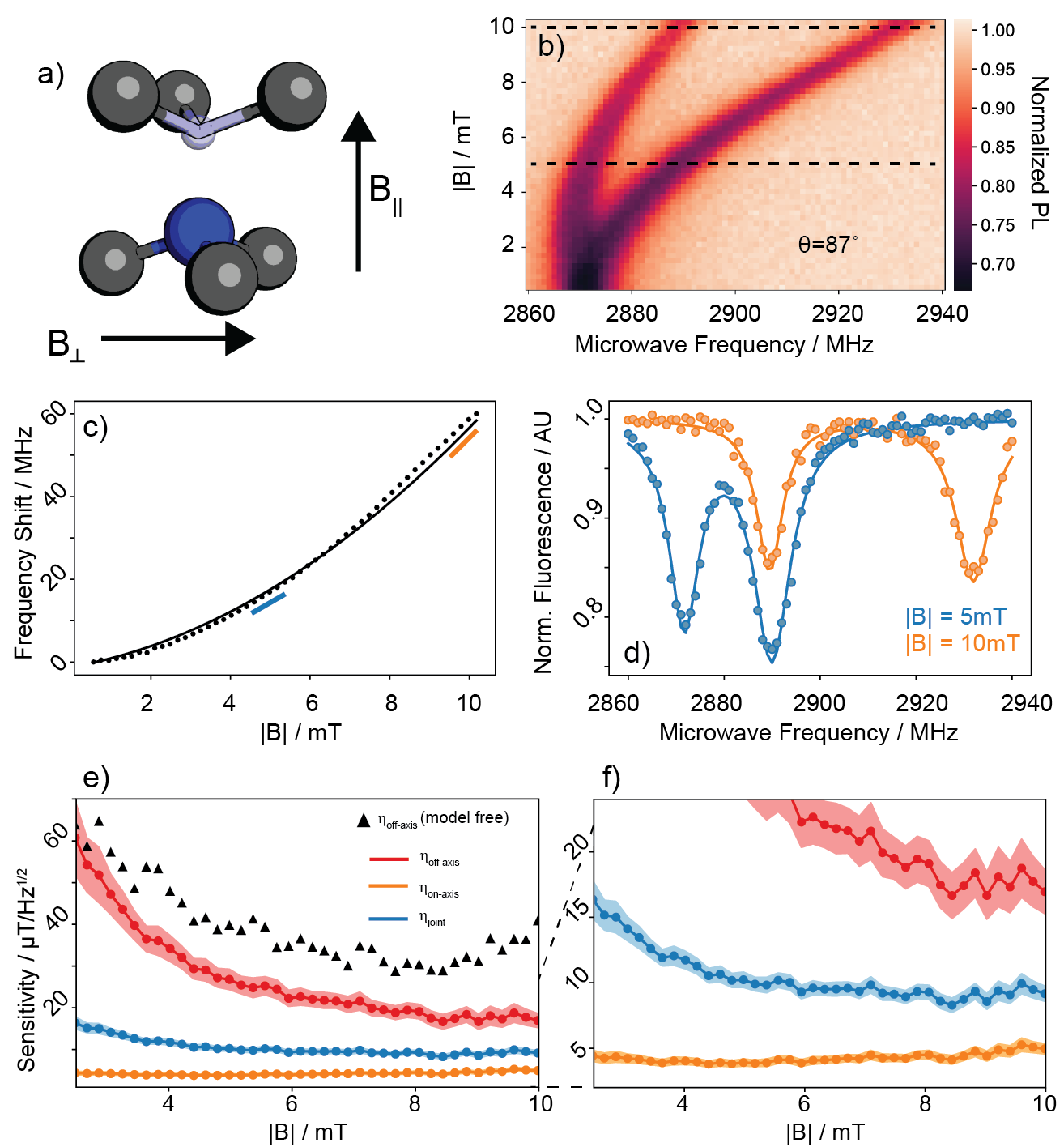}
\caption{\label{fig:schematic} (a) Schematic of the NV center, illustrating the convention used to label the magnetic field directions. (b) Optically-detected magnetic resonance (ODMR) of a single NV center as a function of applied magnetic field. The field is applied at an angle of $\approx 87^{\circ}$ with respect to the zero-field quantization axis. (c) Experimentally determined frequency shift of the higher energy transition as a function of external field (dots). Overlaid is the scaling expected from the perturbation theory description in Eq. \ref{eq:pt}. Colored lines highlight the increase in slope as a function of magnetic field. (d) ODMR spectra for two different magnetic fields (dashed lines in (b)) demonstrating the decrease in contrast as transverse field increases. (e) The experimentally-determined sensitivities for on-axis and off-axis fields, as well as the joint sensitivity (defined in the main text). The sensitivity to off-axis fields improves as the transverse bias field is increased. Shaded regions indicate 95\% confidence intervals from fit parameters. (f) Zoom-in of the sensitivities to show the behavior of the on-axis and joint sensitivity.}
\end{figure}

NV centers are representative of a broad class of point defects where the energy levels and dynamics of the spin states can be detected through changes in the optical properties\cite{chatterjeeSemiconductorQubitsPractice2021,awschalomQuantumTechnologiesOptically2018}. Specifically, in the case of NV centers, room-temperature optically detected magnetic resonance (ODMR) can be realized by monitoring the change in photoluminescence in response to microwave driving fields. This allows  the resonant frequencies of the spin transitions to be determined. The NV center has a ground state with $S=1$ and a zero-field splitting (ZFS) of $D=2\pi \times 2870\,MHz$, which defines a natural quantization axis at low magnetic fields. Two spin transitions are observed, corresponding to transitions from the $m_s = 0$ state to states with $m_s = \pm 1$ character. Neglecting hyperfine effects, the Hamiltonian for this system can be written as:
\begin{equation}
    \frac{H}{\hbar} = D S_z^2 + \gamma_e (B_x S_x + B_y S_y + B_z S_z),
\end{equation}
where $\gamma_e = 2\pi\times 28\,GHz/T$ is the gyromagnetic ratio, $B_i$ is the $i^{th}$ component of the magnetic field, and $S_i$ is the $i^{th}$ spin operator ($i=x,y,z$) in the NV center reference frame. We note that Hamiltonians of this form are ubiquitous in defect-based quantum sensors, meaning our approach can be broadly applied to a wide range of defect systems.

The NV center is typically used to sense the magnetic field component along the $z$ direction (aligned with the ZFS axis, which we label $B_\parallel$ here) since this Zeeman term commutes with the zero-field Hamiltonian. This gives rise to frequency shifts while avoiding the mixing of eigenstates which leads to ODMR contrast reduction. The linear Zeeman shift of the transition energies is:
\begin{equation}
    \omega_\pm = D \pm \gamma_e B_\parallel
\end{equation}

Additionally, however, magnetic fields transverse to the ZFS axis can give rise to frequency shifts. For moderate ($<10\,mT$) fields, these are much smaller than the linear shift and typically neglected. A perturbation theory treatment for the transition frequency shift resulting from these fields yields (to second order)\cite{welterScanningNitrogenvacancyCenter2022}:
\begin{equation}
    \omega_\pm = D \pm \gamma_e B_\parallel + \frac{\gamma_e^2 B_\perp^2}{D\pm \gamma_e B_\parallel} + \frac{1}{2}\frac{\gamma_e^2 B_\perp^2}{D\mp \gamma_e B_\parallel}
\label{eq:pt}
\end{equation}
where $B_\perp$ is the magnetic field component which lies in a plane normal to the quantization axis. Physically, $B_\parallel$ ($B_\perp$) refers to the magnetic field component which is parallel to (normal to) the N-V bond axis. The lab-frame direction will depend on diamond orientation and mounting. 

Thus, the effect of the transverse field is on the order of $\approx \frac{3}{2}\frac{\gamma_e^2 B_\perp^2}{D}$\cite{welterScanningNitrogenvacancyCenter2022}, and shifts both transitions to higher energy. This, in principle, allows both $B_\parallel$ and $B_\perp$ to be determined from a single ODMR spectrum. In practice, however, the large value of the ZFS for the NV center has the effect of suppressing frequency shifts from small transverse fields. A field of $100\mu T$ in the parallel direction results in a frequency shift of $2.8\,MHz$, while the frequency shift for a transverse field of the same magnitude is only $\approx 4\,kHz$. Figure \ref{fig:schematic} shows ODMR data for a single NV center in response to magnetic field applied at an angle of $\theta\approx87^{\circ}$ with respect to the ZFS axis. These data demonstrate both the nonlinear frequency shift from the off-axis fields and the decrease in ODMR contrast at higher fields resulting from the mixing of eigenstates. Thus, directly measuring both field components has been restricted to a limited number of systems where the magnetic field is large\cite{dovzhenkoMagnetostaticTwistsRoomtemperature2018a}, but not too large\cite{welterScanningNitrogenvacancyCenter2022}. We demonstrate here, however, that careful engineering of the static magnetic field used to split the $m_s \pm 1$ transitions (the ``bias'' field) allows us to leverage the nonlinear shift from the transverse magnetic fields, extendable to other optically-addressable spin qubit systems. 

Aiming to sense the component of some small target field transverse to the ZFS axis ($B_\perp ^{(t)}$), we introduce a moderate ($<10m T$) bias field in the transverse direction ($B_\perp ^{(b)}$). Along the direction of the applied field, the $B_\perp^2$ dependence in Eq \ref{eq:pt} can then be written as:
\begin{equation}
    B_\perp ^2 = {B_\perp ^{(b)}}^2 + 2B_\perp ^{(b)}B_\perp ^{(t)} + {B_\perp ^{(t)}}^2.
\end{equation}
For small target fields, we can neglect the ${B_\perp ^{(t)}}^2$ term, leaving a \textit{linear} dependence on the target field with no ambiguity about the sign of the field, and a prefactor which depends on the magnitude of the static bias field, $2B_\perp ^{(b)}B_\perp ^{(t)}$. We draw an analogy between optical heterodyne detection - where a reference optical field is used to mitigate the phase-insensitivity and nonlinear response of square-law optical detectors - with this approach of optimizing the transverse bias field. Transverse fields orthogonal to the applied bias field are not enhanced in this way; since we consider small fields we neglect the effect of these signals, \textit{i.e.} we selectively enhance transeverse fields components along the direction of the applied bias field. By changing the direction of the applied bias field, different components of the transverse field can be selectively enhanced, providing a route towards vector magnetometry of off-axis fields. By utilizing static bias fields with significant transverse components, we can improve the sensitivity of our NV center to small transverse target fields. Importantly, using the two peaks in the ODMR spectrum enables both the parallel and transverse components of the magnetic field to be measured simultaneously. 

Figure \ref{fig:schematic}(e) shows our experimentally determined sensitivity as a function of the applied transverse field. We determine this in two ways; first we fit the ODMR spectra at each field to estimate the signal-to-noise and maximum slope of each peak, and from this the minimum frequency shift. We also numerically calculate the derivative of the peak position with respect to magnetic field to determine the minimum detectable change in transverse field. Second, we use the linewidth and contrast determined from the fit to calculate the theoretical sensitivity from an analytical model (Eq. \ref{eq:sens}). Both approaches are in good agreement; the empirical approach (which does not assume a particular noise source, labelled ``model free'' in Figure \ref{fig:schematic}) is within a factor of 1.5 of the shot-noise limited model. In these experiments, we are able to reach sensitivities to transverse magnetic fields of $\eta_{\perp} \approx 30\,\mu T / \sqrt{Hz}$

\begin{figure}
\includegraphics[width=0.45\textwidth]{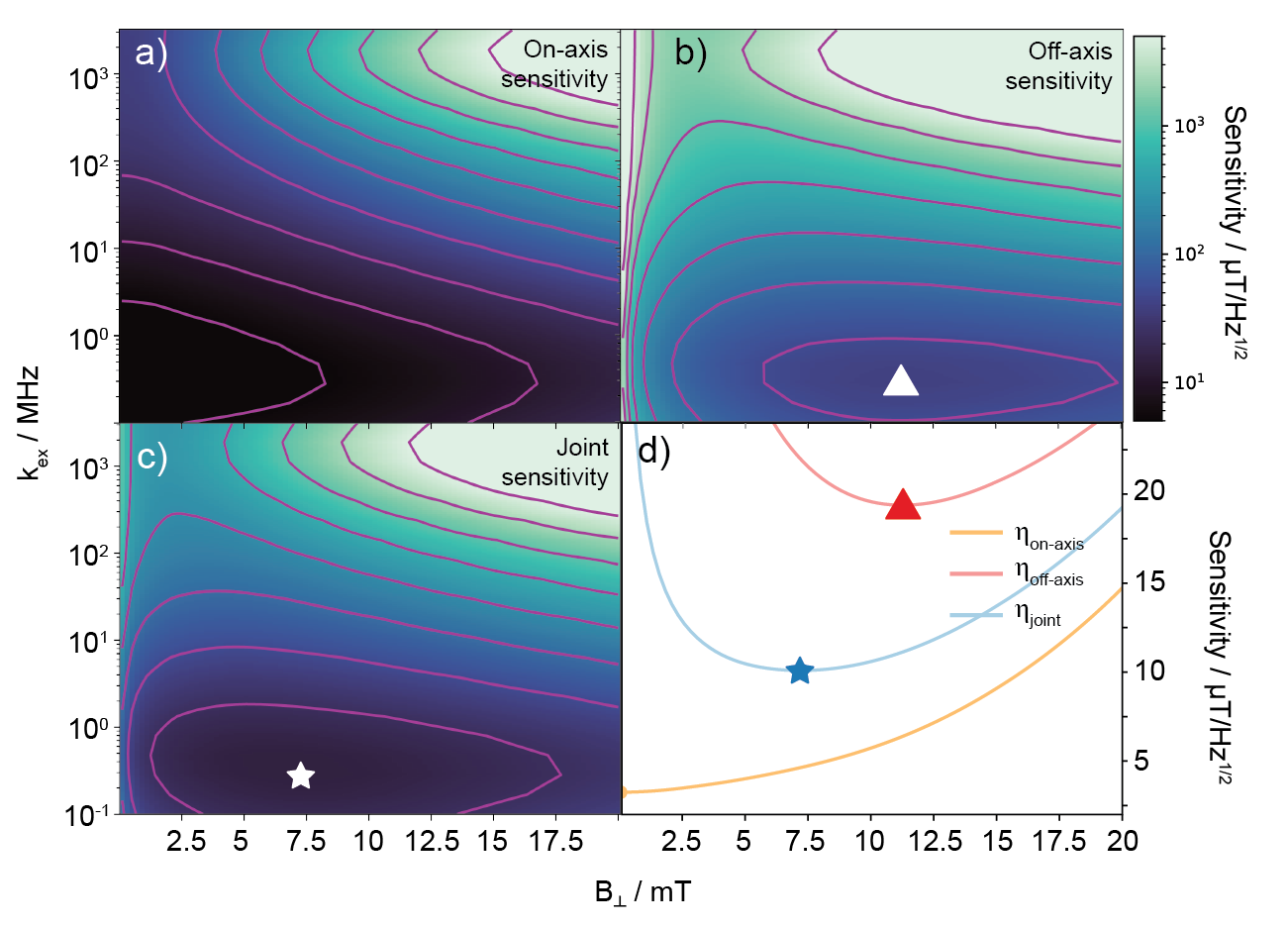}
\caption{\label{fig:sens} Simulated sensitivities as a function of transverse bias field and optical excitation rate for (a) sensing parallel fields, $\eta_\parallel$, (b) sensing transverse fields, $\eta_\perp$, (c) jointly-optimized sensing, $\sqrt{\eta_\parallel \eta_\perp}$. Countour lines are equally spaced on a logarithmic scale. (d) 1D plots of the minimum $\eta$ at each magnetic field for the parallel, perpendicular, and joint sensitivities.}
\end{figure}

Arbitrarily increasing the transverse bias field, however, does not yield optimal sensitivity; both the steady-state photoluminescence and the ODMR contrast decrease under moderate transverse bias fields, with the contrast dropping to almost zero at larger fields\cite{welterScanningNitrogenvacancyCenter2022,tetienneMagneticfielddependentPhotodynamicsSingle2012}. To estimate the optimal sensitivity and limitations of the NV center, we numerically simulate the contrast and photoluminescence as a function of transverse magnetic field and optical excitation rate (\textit{i.e.} laser power) by finding the steady-state behavior of the density matrix, $\rho$. We numerically solve the Lindblad equation using QuTiP\cite{johanssonQuTiPOpensourcePython2012,johanssonQuTiPPythonFramework2013}:
\begin{equation}
    \frac{d \rho (t)}{d t} = -\frac{i}{\hbar}[H,\rho] - \frac{1}{2}\sum_{k=0}^m(L_k ^\dag L_k \rho + \rho L_k ^\dag L_k) + \sum_{k=0}^m L_k \rho L_k ^\dag,
\end{equation}
where the Kraus operators ($L_k$) are of the form described by Stefan \textit{et al} \cite{stefanMultiangleReconstructionDomain2021}. We use the photophysical rates from Tetienne \textit{et al} \cite{tetienneMagneticfielddependentPhotodynamicsSingle2012} for the simulations, and model laser- and microwave-power dependent broadening of the spin transition using the expressions from Dreau \textit{et al}\cite{dreauAvoidingPowerBroadening2011}. Specifically, the processes we consider are the radiative relaxation (which links ground and excited states of the same spin projection with rate $\gamma_r$), intersystem crossing from the excited states to a common shelving state (defined by spin-dependent rates $\gamma_{ISC,0}$ and $\gamma_{ISC,\pm1}$), and relaxation of the shelving state to the ground state (assumed to be spin projection-independent\cite{stefanMultiangleReconstructionDomain2021}, with rate $\gamma_{SS}$). We then estimate the change in photoluminescence from changes in the steady state population of the excited states of the NV center.  Parameters used in the simulations are given in Table \ref{tab:rates}.

\begin{table}[]
    \centering
    \begin{tabular}{|c|c|} \hline
    Parameter  &  Value \\ \hline \hline
        D  &  $2\pi \times 2870$ MHz \\ \hline
        $\Omega$   & $2 \pi \times 1$ MHz\\ \hline
        $\gamma_r$    &  65 MHz \\ \hline
        $\gamma_{ISC,0}$ &  11 MHz\\ \hline
        $\gamma_{ISC,\pm1}$  &  80 MHz\\ \hline
        $\gamma_{SS}$   &  3 MHz\\ \hline
        
    \end{tabular}
\caption{Parameters used for the numerical simulations described in the text. $\gamma_r$ is the radiative rate, $\gamma_{ISC,m}$ is the rate of intersystem crossing to the shelving state from the excited state with spin projection number $m$, $\gamma_{SS}$ is the relaxation from the shelving state, and $\Omega$ is the Rabi frequency of the microwave drive.}
\label{tab:rates}
    
\end{table}

From these simulations, we calculate the sensitivity of the NV center to both parallel ($\eta_\parallel$) and transverse fields ($\eta_\perp$) using\cite{taylorHighsensitivityDiamondMagnetometer2008} $\eta = \delta B \sqrt{T}$ . In CW ODMR experiments, the sensitivity\cite{dreauAvoidingPowerBroadening2011} is:
\begin{equation}
    \eta_\parallel = P_F \frac{\Delta \nu}{C \sqrt{R}}\frac{1}{\gamma_e},
\end{equation}
where $\Delta \nu$ is the peak full-width at half-maximum, $C$ is the ODMR peak contrast, $R$
is the photon detection rate, and $P_F$ is a prefactor related to the lineshape ($\frac{4}{3\sqrt{3}}$ for a Lorentzian peak). For the transverse component, we can use Equation \ref{eq:pt} to determine an analogous expression in the limit of small $B_\perp^{(t)}$ and fixed $B_\perp^{(b)}$:
 \begin{equation}
    \eta_\perp = P_F \frac{\Delta \nu}{C \sqrt{R}}\frac{D}{3\gamma_e^2 B_\perp^{(b)}}.
    \label{eq:sens}
\end{equation}
Note, the contrast is a function of the transverse field, decreasing as the magnitude of the field increases. 

The calculated sensitivities as a function of both laser power (excitation rate) and transverse bias field are shown in Figure \ref{fig:sens}. The sensitivity of the NV center to transverse field components can be improved with larger static bias fields, up to the point where the decrease in contrast begins to degrade sensitivity ($\approx 10\,mT$). The sensitivity to magnetic fields parallel to the ZFS axis, however, is only degraded by the application of the transverse bias field because of the decreasing ODMR contrast. To simultaneously sense both components, we define a joint sensitivity $\eta_{joint} = \sqrt{\eta_\parallel \eta_\perp}$, shown in Figure \ref{fig:sens}c. This defines an optimum range of bias fields to sense both parallel and transverse field components; we identify bias fields in the range of $5\,mT$ as optimal, noting that the joint sensitivity is rather slowly varying in the region around this point. These simulated curves agree well with our experimentally-defined sensitivities in Figure \ref{fig:schematic}(e). While the absolute values of the sensitivity depend on material properties (linewidth, contrast, detected count rate), we emphasize that a \textit{relative} improvement in sensitivity can be realized with this approach even in the limit of broad linewidths or low contrast. Our analysis highlights the region where the frequency shift from transverse fields is increased with minimal degradation in contrast arising from mixing of the eigenstates; this can be applied to a wide range of NV center environments. 

\begin{figure}
\includegraphics[width=0.45\textwidth]{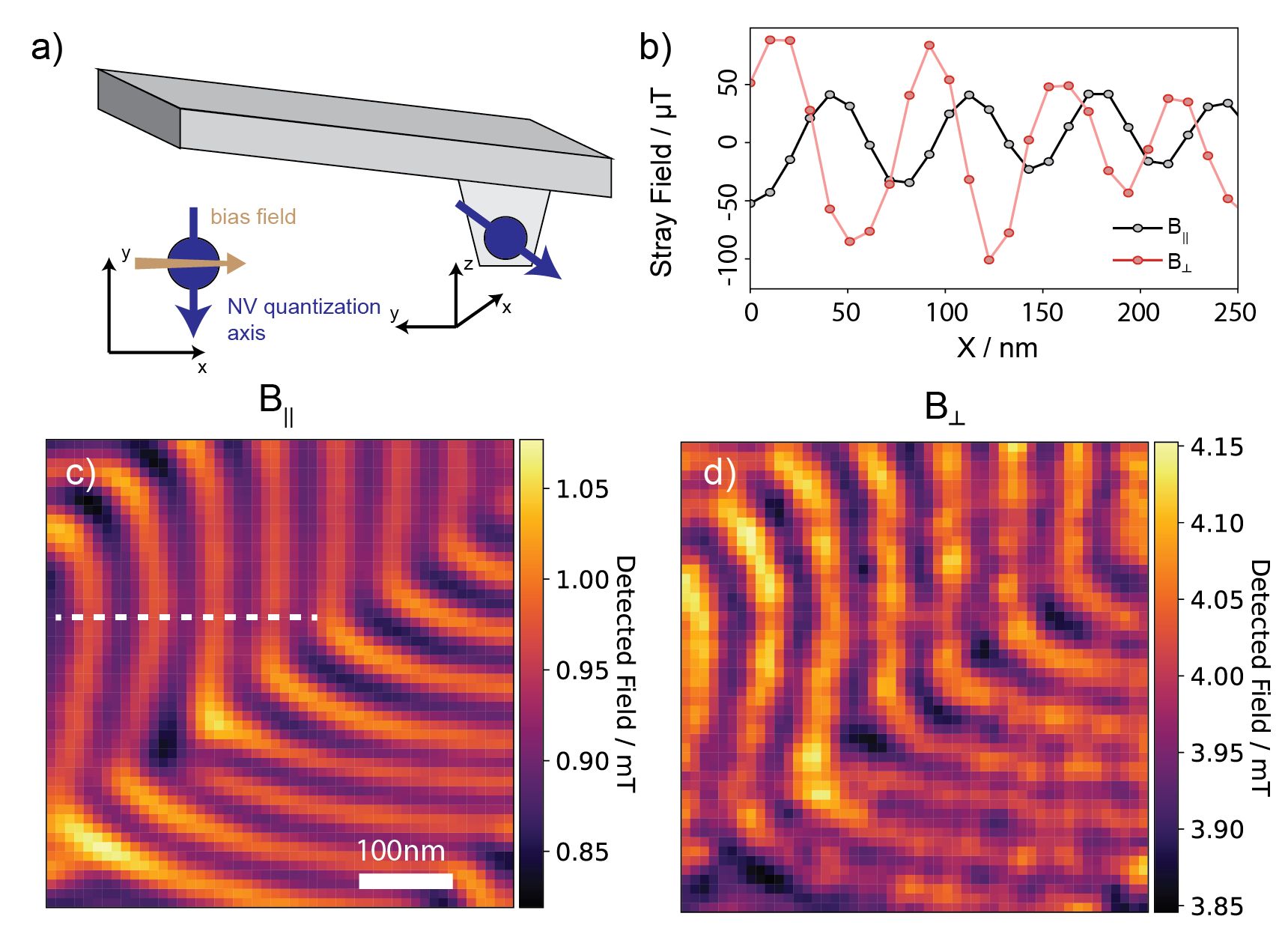}
\caption{\label{fig:bfo} (a) Schematic of the scanning tip geometry and magnetic fields. Permanent magnets are used to apply a large bias field primarily (but not entirely) in the $x-z$ plane. The determined components of this field are $B_{||}=0.9 \,mT$ $B_{\perp}=4.0\,mT$. (b) Linecuts along the dashed line (panel (c) ) showing the parallel and transverse components of the fields from the BiFeO$_3$ sample with the static field subtracted. 2D spatial maps of the (c) magnetic field parallel to the NV ZFS axis, and (d) transverse to the NV ZFS axis.}
\end{figure}

Using experimental parameters from our scanning NV instrument (Qnami Quantilever MX+, $R = 4\times10^5$ counts-per-second, $\Delta \nu = 10\,MHz$, $C=0.15$), we can estimate a sensitivity to small transverse fields of $\eta_\perp \approx 20\,\mu T / \sqrt{Hz}$ for a $5\,mT$ bias (similar to the sample used for the data in Figure \ref{eq:sens}). For comparison, with a weak bias field of $B_\perp^{(b)} = 0.1\,mT$, the sensitivity is instead of the order $\eta_\perp \approx 1 mT / \sqrt{Hz}$.

Demonstrating the improved transverse sensitivity using our bias field optimization, we image the vector components of the spin cycloid structure in BiFeO$_3$, a canted antiferromagnet with a spatially-varying magnetization that yields periodic stray fields of $\lambda\approx 65\,$nm\cite{burnsExperimentalistGuideCycloid2020}. The combination of fine spatial features with well-characterized, small magnitude magnetic fields ($\approx 100 \mu T$) makes this an exacting test of the capabilities and limitations of our combined parallel-transverse sensing approach. Our objective here is to demonstrate that we can use the peak shift from both ODMR peaks to extract the stray field generated by the cycloid along multiple axes simultaneously, though we emphasize that our bias field engineering approach can be used for single-point measurements also.

Figure \ref{fig:bfo} shows data collected on a 100 nm, $(001)$ oriented thin film of BiFeO$_3$ deposited on a $(110)$-oriented DyScO$_3$ substrate (sample details are given fully in Reference \cite{meisenheimerSwitchingSpinCycloid2024}) taken with a scanning-tip NV microscope (Qnami ProteusQ). We apply a bias field using a permanent magnet at an angle of $\approx 75^{\circ}$ with respect to the NV center ZFS axis (shown in Figure \ref{fig:bfo}, calculated as the dot product between the quantization axis and the applied field). At each point in space, we measure a full ODMR spectrum including both transitions to simultaneously extract the parallel and transverse components of the stray field. We emphasize that no modifications from the standard scanning NV experiment were required beyond application of a moderate external bias in an in-plane direction, achieved by mounting a permanent magnet $\approx$1" from the sample. The spatial images show the characteristic out-of-phase sinusoidal relationship between orthogonal vector components of the magnetic field, with stray field components from the cycloid on the order of $\pm 100 \mu T$, consistent with previous observations of this system\cite{meisenheimerSwitchingSpinCycloid2024, haykalAntiferromagneticTexturesBiFeO32020}. The transverse fields we detect here are 100$\times$ smaller than those previously imaged without bias-field optimization.\cite{dovzhenkoMagnetostaticTwistsRoomtemperature2018a}. Though reconstructing the vector magnetic field from these spatial images is possible without the bias-field engineering described here, the process is prone to artifacts from finite-size effects. These can, however, be mitigated by measuring additional vector components \cite{broadwayImprovedCurrentDensity2020}, thus, the ability to measure two orthogonal components of the stray field has significant value even in this context. 

We further demonstrate the utility of our bias field engineering approach by measuring a 4-point ``iso-B'' image of the vector fields with our scanning-tip system. Iso-B measurements measure the photoluminescence in response to one or two microwave fields\cite{rondinNanoscaleMagneticField2012,rondinStrayfieldImagingMagnetic2013}, enabling the magnetic field texture (or, quantitatively, field contours) of a sample to be imaged without the time-consuming step of acquiring an entire ODMR spectrum at each pixel. However, this precludes simple reconstruction of the vector magnetic field with the usual approaches; here, we extend this approach to measure the photoluminescence at four microwave drives (two for each peak) enabling us to extract the magnetic field contours of the BiFeO$_3$ cycloid for only twice the acquisition time of a typical iso-B measurement. 

\begin{figure}
\includegraphics[width=0.45\textwidth]{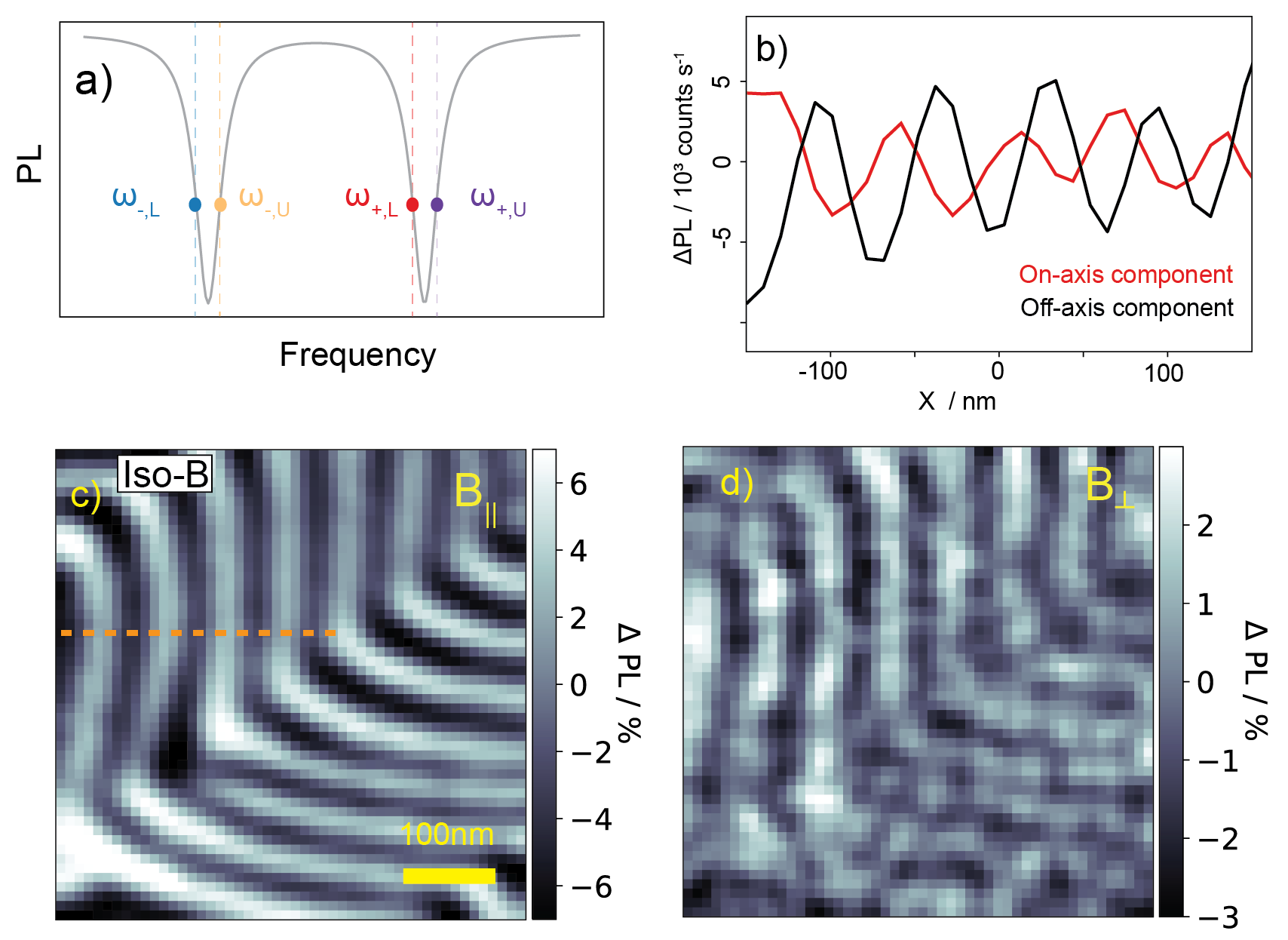}\caption{\label{fig:bfo_isob} (a) The four-point iso-B method monitors two frequency points per peak to extract both parallel and transverse magnetic field textures (b) Iso-B image plotting the relative change in PL of the magnetic field components parallel to the ZFS axis. The heatmap indicates the relative change in detected photon count rate. (c) Iso-B image of the magnetic field components transverse to the ZFS axis (d) Linecuts of the two components along the dashed line in (b), again highlighting the expected relationship for the stray fields in BiFeO$_3$.}
\end{figure}

\begin{table}
    \centering
    \begin{tabular}{|c|c|c|} \hline 
       System  &  D / $rad\,s^{-1}$ & Reference \\ \hline
       hBN & $2\pi \times 3480$ & \cite{ramsayCoherenceProtectionSpin2023}\\ \hline
        NV (diamond) & $2\pi \times 2870$ & \cite{chatterjeeSemiconductorQubitsPractice2021}\\ \hline 
        TR12 (diamond) & $2\pi \times 1636$ & \cite{foglszingerTR12CentersDiamond2022}\\ \hline
        SiV$^{0}$ (diamond) & $2\pi \times 944$ & \cite{roseObservationEnvironmentallyInsensitive2018}\\ \hline 
        VV (SiC) & $2\pi \times 1300$ & \cite{sonDevelopingSiliconCarbide2020}\\ \hline 
    \end{tabular}
    \caption{Other known point defects with $S=1$ and their zero-field splittings (D). We use these values to simulate the gain in sensitivity to transverse fields in the main text.}
    \label{tab:defects}
\end{table}

In the 4-point iso-B measurement, we track two microwave frequencies (``lower'', L, and ``upper'', U) for each of the two peaks (labelled $+$, $-$), as illustrated in Figure \ref{fig:bfo_isob}. The frequencies are chosen to lie on the maximum slope of the peaks, which for a Lorentzian lineshape occurs at $\omega = \pm \frac{1}{2}\Gamma/\sqrt{3}$. We calculate the iso-B signal arising from the parallel and transverse components as: 
\begin{align}
\Delta_\parallel & = \frac{1}{2}\bigg(PL(\omega_{+,U}) - PL(\omega_{+,L}) + PL(\omega_{-,U}) - PL(\omega_{-,L}) \bigg) \\
\Delta_\perp & = \frac{1}{2}\bigg(PL(\omega_{m,U}) - PL(\omega_{m,L}) - PL(\omega_{+,U}) + PL(\omega_{+,L}) \bigg).
\end{align}
With this approach, both the parallel and transverse magnetic textures from the BiFeO$_3$ cycloid can be recovered, shown in Figure \ref{fig:bfo_isob}b,c, directly comparable to Figure \ref{fig:bfo}c,d. This approach greatly improves the data acquisition rate; the data shown in Figure \ref{fig:bfo_isob}b,c are collected with an integration time of 60ms per frequency, per point, enabling the entire dataset to be acquired in 10 minutes. In contrast, the conventional approach to vector magnetometry which acquires the full ODMR spectrum at each point (Figure \ref{fig:bfo}c), takes $\approx$2 hours. The 4-point iso-B method enabled by engineering the bias field thus represents an order-of-magnitude speed-up for imaging multiple magnetic field components simultaneously.

\begin{figure}
\includegraphics[width=0.45\textwidth]{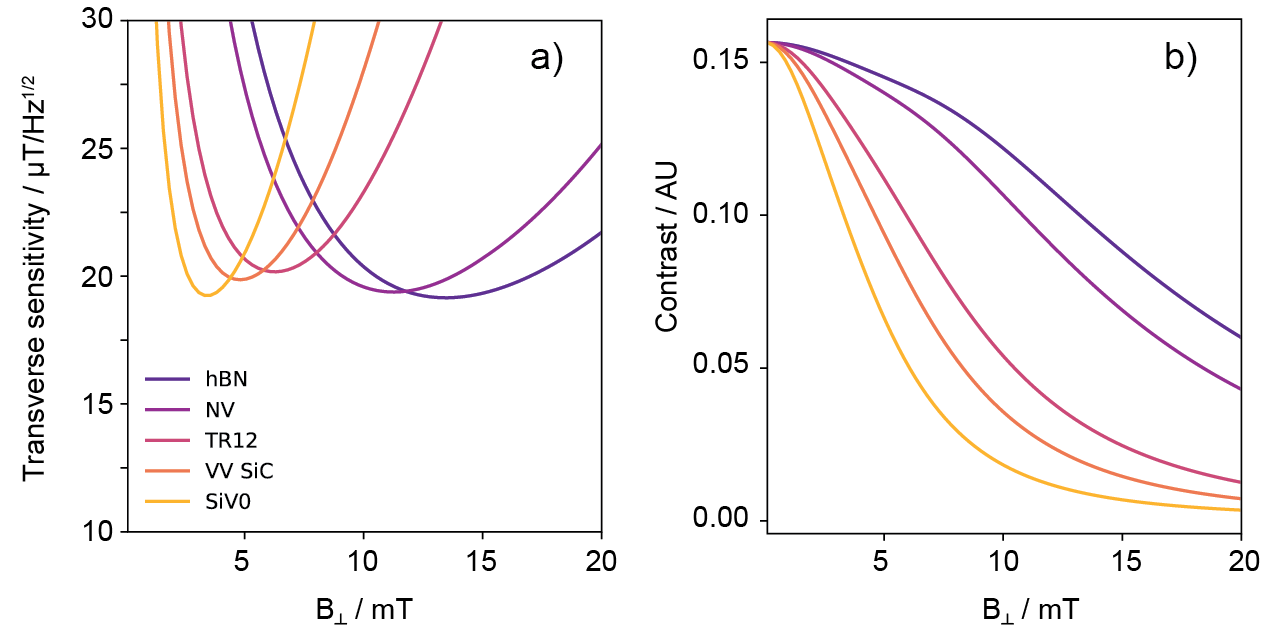}
\caption{\label{fig:defect_sims} (a) The transverse sensitivity $\eta_\perp$ for a range of defect zero-field splittings. (b) The variation in contrast as a function of transverse field for the various defect Hamiltonians.}
\end{figure}

Finally, we consider the applicability of this approach to systems beyond the NV center in diamond. A vast number of point defects with optically-addressable spin states have been detailed; here we focus on $S=1$ systems to facilitate comparison with the NV center. Example alternative systems are listed in Table \ref{tab:defects}. The photophysics of each of these systems may differ substantially, leading to greater or less contrast, spin polarization, photoluminescence \textit{etc}. To simplify the comparison, we assume all photophysics rates are the same as the NV center model used (Table \ref{tab:rates}), and focus on changes in the transverse sensitivity due to differences in mixing of the spin states. We observe that the optimum bias field for transverse field sensing decreases with decreasing $D$, but the optimum sensitivity is only weakly dependent on the bias field magnitude. This comes from two competing effects; 1) the suppression of the second order frequency shift in Equations \ref{eq:pt} and \ref{eq:sens} is less as the $D$ parameter decreases, however, 2) this also has the effect of reducing the contrast for smaller bias fields, as shown in Figure \ref{fig:defect_sims}. Thus, while the specific optimum operating point may vary between systems, we expect the approach we develop here to be broadly applicable to vector magnetometry with point defects.

In summary, we demonstrate that single point defects can be used to sense multiple components of a magnetic field simultaneously through the application of a static bias field orthogonal to the zero-field quantization axis. Simulating the photophysics of this system allows us to determine an optimal operating point to sense both parallel and transverse fields, enabling orders-of-magnitude improvement in the transverse sensitivity with minimal degradation of the sensitivity to parallel fields. We validate this approach experimentally by imaging the stray fields generated by the spin cycloid in BiFeO$_3$ with conventional ODMR and demonstrate a 4-point estimation method to increase data collection rates.  This approach can be broadly applied to a range of systems and could be extended to enable the detection of small transverse AC magnetometry signals.

\section*{Supplementary Material}
Additional sample details and experimental methods, description of the experimental sensitivity estimations, and derivation of the off-axis sensitivity equation are included in the Supplementary Material.

\section*{Acknowledgements}
P.S. acknowledges support from the Massachusetts Technology Collaborative, Award number \#22032. R.R. acknowledges funding from the Army Research Office under the ETHOS MURI via cooperative agreement W911NF-21-2-0162.

\section*{References}

\bibliography{OffAxis}

\end{document}